\documentclass[amsmath, amsfonts, superscriptaddress, showpacs, twocolumn, prl]{revtex4}
\usepackage{graphicx}
\usepackage{epsfig}
\usepackage{bm}
\usepackage{dcolumn}
\usepackage{amsmath}
\usepackage{amssymb}
\usepackage{color}

\DeclareFontFamily{U}{wncy}{}
\DeclareFontShape{U}{wncy}{m}{n}{<->wncyr10}{}
\DeclareSymbolFont{mcy}{U}{wncy}{m}{n}
\DeclareMathSymbol{\Sh}{\mathord}{mcy}{"58}
\DeclareMathSymbol{\Ka}{\mathord}{mcy}{"4B}

\newcommand{\be}{\begin{equation}}
\newcommand{\ee}{\end{equation}}

\begin{document}

\title{Dark counts initiated by macroscopic quantum tunneling in NbN superconducting photon detectors}

\author{A.~Murphy}
\affiliation{Department of Physics, University of Illinois at
Urbana-Champaign, Urbana, Illinois 61801, USA}

\author{A.~Semenov}
\affiliation{Moscow State Pedagogical University, 1 Malaya Pirogovskaya, 119991 Moscow, Russia}
\affiliation{Moscow Institute of Physics and Technology, 141700, Dolgoprudny, Moscow Region, Russia}

\author{A.~Korneev}
\affiliation{Moscow State Pedagogical University, 1 Malaya Pirogovskaya, 119991 Moscow, Russia}
\affiliation{Moscow Institute of Physics and Technology, 141700, Dolgoprudny, Moscow Region, Russia}

\author{Yu. Korneeva}
\affiliation{Moscow State Pedagogical University, 1 Malaya Pirogovskaya, 119991 Moscow, Russia}

\author{G.~Gol'tsman}
\affiliation{Moscow State Pedagogical University, 1 Malaya Pirogovskaya, 119991 Moscow, Russia}
\affiliation{National Research University Higher School of Economics, Moscow 101000, Russia}

\author{A.~Bezryadin} 	
\affiliation{Department of Physics, University of Illinois at
Urbana-Champaign, Urbana, Illinois 61801, USA}

\begin{abstract}
We perform measurements of the switching current distributions of three $w$ $\approx$ 120 nm
wide, 4 nm thick NbN superconducting strips which are used for single-photon detectors. These  strips are much wider than the diameter the vortex cores, so they are classified as quasi-two-dimensional (quasi-2D). We discover evidence of macroscopic quantum tunneling by observing the saturation of the standard deviation of the switching distributions at temperatures around 2 K. We analyze our results using the Kurkij\"arvi-Garg model and find that the escape temperature also saturates at low temperatures, confirming that at sufficiently low temperatures, macroscopic quantum tunneling is possible in quasi-2D strips and can contribute to dark counts observed in single photon detectors.
\end{abstract}

\date{October 27, 2014}


\maketitle

\section{I. Introduction}

Quantum tunneling between macroscopically distinct states has been studied extensively in Josephson junctions and nanowires \cite{Sahu-Nature, Clarke-Science}. Some of the most basic evidence for macroscopic quantum tunneling (MQT) in these systems comes from the saturation of the standard deviation of switching current distributions and the saturation of escape temperature at bath temperatures below some threshold value \cite{Tinkham, Aref-PRB, Clarke-Science, Jackel-PRL, Voss-PRL}. Exploring quantum tunneling in quasi-2D superconductors has recently become a particularly interesting topic as the practical use of quasi-2D NbN strips as single photon detectors has grown \cite{Reithmaier-SR, Gol'tsman-APL, Natarajan-SST}. In practice, these detectors are driven at currents near their critical currents so that when a photon strikes the superconductor, it can cause a segment of the strip to become normal for a short period of time, registering a voltage pulse. However there also exists a rate of false events, known as dark counts, whereby a voltage pulse can be detected without an incident photon. A possible origin of these dark counts is thermally activated escape from the superconducting state \cite{Bartolf-PRB, Yamashita-APL, Kitaygorsky-2007, Tafuri-EPL, Bulaevskii-PRB, Engel-Physica}. Such thermal dark counts can be suppressed by reducing temperature. A second possible origin of dark counts is MQT of vortices or phase slips between the ends of the quasi-2D strip \cite{Kitaygorsky-2005}. This option has been speculated theoretically but not yet observed experimentally. The rate of MQT is not expected to be strongly affected by temperature, and therefore MQT could provide a base-level dark count rate present below some threshold quantum temperature. Note that we understand a phase-slip event to be any process which leads to a quantized phase change of the order parameter by $2 \pi$ between the ends of the strip, which represents the elementary dissipative event in the superconductor.

As dark counts in wide strips have been investigated, a debate over the microscopic process by which the strips switch from the superconducting state to the normal state has emerged. Recent theories and experiments have supported three different escape processes (all of which are $2 \pi$ phase-slip events), including single vortices crossing an edge barrier \cite{Bartolf-PRB, Qui-PRB, Tafuri-EPL, Bulaevskii-PRB}, vortex-antivortex pairs splitting apart under the action of the Lorentz force \cite{Bartolf-PRB, Yamashita-APL, Kitaygorsky-2007, Engel-Physica} and escape through an energy saddle point which does not involve a vortex core \cite{Vodolazov-PRB, Bell-PRB, Kitaygorsky-2005}. Such a saddle point, with a suppressed but still above zero order parameter, has been used to explain Little-type phase-slips in one-dimensional superconducting wires \cite{Little-PR, Langer-PR}. Recently, it was generalized theoretically to quasi-2D superconducting strips \cite{Vodolazov-PRB}. After crossing the energy saddle point, i.e. when the free energy is already lowered, a vortex is formed in the strip. The Lorentz force then pushes the vortex across the strip causing a phase-slip. So in any theory, it is expected that a vortex or vortex-antivortex pair is needed to produce a phase-slip. Yet in the vortex-free saddle point model, the maximum of the free energy is achieved before the vortex core is created. For the case of MQT, this model is most plausible because it doesn't involve quasiparticle dissipation. Note that according to Caldeira-Leggett theory any dissipation reduces the rate of quantum tunneling exponentially \cite{Caldeira-AP}.

Typically, experimental analysis of dark counts focuses on fitting dark count rate data to one or more of these three theories. Most experiments so far were performed in the range of intermediate temperatures, where superconducting single-photon detectors commonly operate and MQT is not expected to be observed. The few experiments which have reached much lower temperatures have drawn conflicting conclusions both on escape method and on presence of MQT \cite{Tafuri-EPL, Yamashita-APL, Kitaygorsky-2005}. It has therefore become apparent that in order to study MQT in wide strips, evidence for MQT should come from new sources such as counting-statistics. 

In this letter, we report evidence of macroscopic quantum tunneling in wide (compared to the size of the vortex core) NbN strips by demonstrating the saturation of the standard deviation of their switching current distributions and the saturation of the escape temperature. Because the switching events are observed without any photon irradiation and with careful multi-stage electromagnetic noise filtering, these events represent so-called "dark counts". We fit our data using the general Kurkij\"arvi-Garg (KG) model and show that the escape rates in such systems can be approximated using forms similar to those derived for Josephson junctions, nanowires and graphene junctions made superconducting by proximity effect \cite{Bezryadin, Kurkijarvi-PRB, Garg-PRB, Aref-PRB, Coskun-PRL}.

\section{II. Experiment}

\begin{figure}[t!]
  \includegraphics[width=8cm]{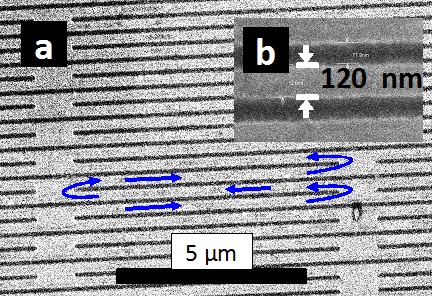}
  \caption{ (a) An SEM image of Sample 4. The NbN wire (lighter region) is roughly 250 $\mu$m in length and about 4 nm thick. The path which the current travels is indicated by arrows. Not shown in the figure, to the far right or the far left, each horizontal NbN segment comes to an abrupt end and has no additional connections to the rest of the superconductor.   (b) A zoomed in view on one segment of the wire showing the width of the wire to be about 120 nm.
 }\label{Fig-sem}\vskip-.5cm
\end{figure}

The samples measured are  approximately 245 $\mu$m long, 4 nm thick and 120 nm wide NbN strips (Fig.~\ref{Fig-sem}). The coherence length $\xi$ and perpendicular magnetic penetration depth $\Lambda$ of similar samples are known to be around 5 nm and 50 $\mu$m respectively \cite{Sindler-SST, Kamlapure-APL}. Because $\xi \ll w$, the strip forming the photon detector is considered quasi-two-dimensional. Yet the supercurrent is expected to be uniform across the width of the strip because $w\ll \Lambda$. The critical temperature of each device is around 10 K (Fig. \ref{Fig-RT}a).

The samples are fabricated from 4 nm thick niobium nitride (NbN) film. The fabrication procedure is similar to the one described in detail in \cite{Goltsman-JMO}. In brief, the NbN film is deposited by DC reactive magnetron sputtering on thermally oxidized silicon wafers. The film is patterned as meander-shaped strip by electron beam lithography in PMMA 950K resist and reactive ion etching in SF$_6$. The strips outside the meander are used for proximity effect correction in e-beam lithography.

Measurements were performed in a He-4 system with a base temperature of 1.45 K. The samples were current-biased through a series resistor of value 46 k$\Omega$; the other end of the NbN strip was connected to the ground through a series resistor of value 1 k$\Omega$. The normal state resistance of our devices $R_n$ was of the order of 1 M$\Omega$. The voltage was measured across the sample directly while the current was determined by measuring the voltage across the 1 k$\Omega$ series resistor. A sinusoidal bias voltage with a corresponding maximum current greater than the critical current of the sample was applied to the system. Signal lines were filtered by pi-filters at high temperature and passed through silver-particle filters at the base temperature. As the bias current increased from zero, the sample suddenly jumped from zero voltage to a non-zero value and the current through the system suddenly decreased. This happens because the sample switches to the resistive state, and the normal state resistance is much larger than the series resistors. The value of current at the moment of the voltage jump, which corresponds to the peak current measured, was recorded as the switching current, $I_{sw}$. This measurement was repeated at least 10,000 times at each temperature for Samples 4 and 12, and 50,000 times for Sample 7. The peak bias voltage was adjusted at each temperature in attempt to keep the sweep speed at the switching current constant with temperature. Note that the current was changing with time as $I(t) =\frac{ V_0}{R_T} sin(\omega t)$ where $V_0$ is the peak bias voltage, $R_T$ is the total resistance of the experimental setup, and $\omega$ = 41 Hz was the frequency. Therefore the sweep speed at $I=I_{sw}$ can be calculated as $\frac{dI}{dt} =\frac{ V_0 \omega}{R_T} \sqrt{1-(I_{sw} R_T/ V_0)^2}$. Sweep speeds at $I_{sw}$ were kept near 3 mA/s, 3.6 mA/s and 2.6 mA/s for Samples 4, 7 and 12 respectively.

\section{III. Analysis}

\begin{figure}[t!]
  \includegraphics[width=8cm]{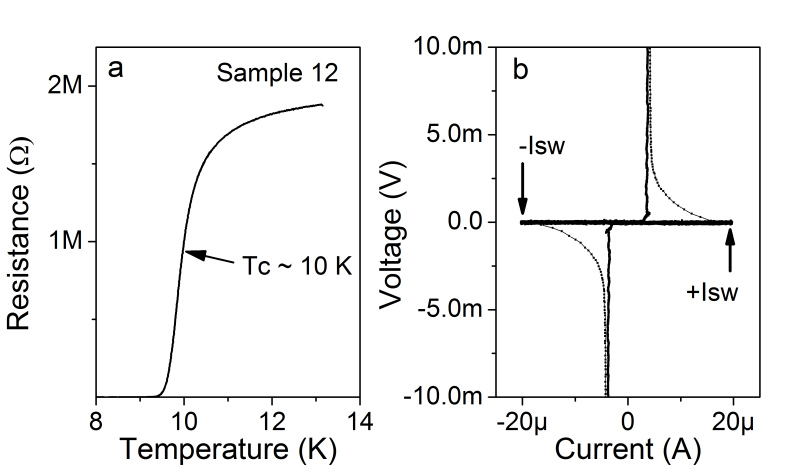}
  \caption{(a) The RT curve of Sample 12 shows a critical temperature around 10 K.
  (b) A typical VI curve. The sample resistance is much larger than the series resistance, thus the current through the sample quickly decreases when the sample becomes resistive.
 }\label{Fig-RT}\vskip-.5cm
\end{figure}

Fig.~\ref{Fig-RT}b shows a typical voltage versus current dependence (VI curve). When a segment of the NbN strip switches from superconducting to normal, the total resistance of the system sharply increases, causing the current through the system to fall. The switching current is marked in Fig.~\ref{Fig-RT}b with arrows. In fact, there are two such switching currents, one occurring at the positive bias ($I_{sw}+$) and one occurring at the negative bias ($I_{sw}-$).

\begin{figure}[t!]
  \includegraphics[width=9cm]{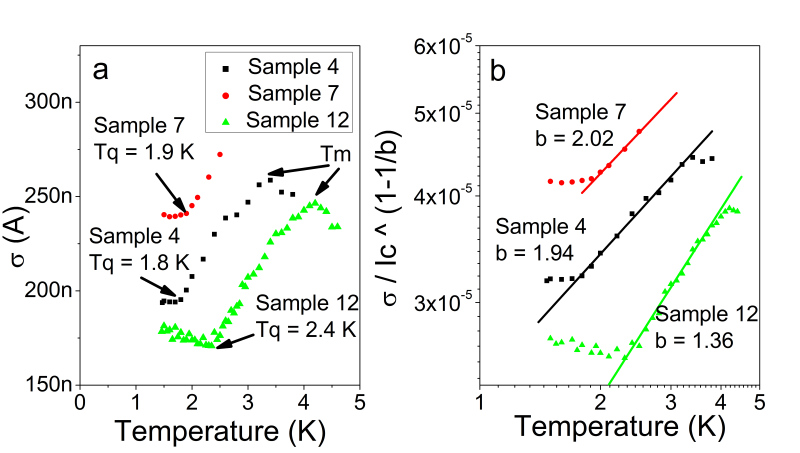}
  \caption{(a) The standard deviation of the switching current distributions of Samples 4, 7 and 12 vs temperature. All three samples show saturation at low temperatures which is known as a signature of macroscopic quantum tunneling. The temperatures at which $\sigma$ saturates are recorded as the quantum temperatures of the strips and are indicated on the plot. For $T > T_m$ multiple phase-slips are necessary for escape. (b) Fits of $\sigma /  I_c^{1-1/b}$ to $T^b$ (solid curves) are performed for $T_q < T < T_m$  in order to find the best-fit power, $b$, for the temperature dependence, according to the Kurkij\"arvi analysis \cite{Kurkijarvi-PRB}. The current dependence in the energy barrier for a phase-slip event has the same power (see Eq. 1). The best-fit powers are shown in the figure and also listed in Table \ref{table-params}. The data and fit for Sample 12 in Figure b are multiplied by a factor of 9 so that all three curves can be easily seen.
 }\label{Fig-sigma}\vskip-.5cm
\end{figure}

At any given temperature, the switching current exhibits a stochastic nature, resulting in a distribution of switching events. The standard deviation $\sigma$ of these distributions, which is plotted versus temperature in Fig.~\ref{Fig-sigma}a, can be analyzed to determine whether quantum or thermal fluctuations are responsible for the strip switching to the resistive state \cite{Kurkijarvi-PRB, Jackel-PRL}. At high temperatures ($T > T_m$ where $T_m$ is the temperature corresponding to the maximum in $\sigma$) standard deviation increases with cooling. At these temperatures, multiple phase slips are necessary to switch the wire to a resistive state \cite{Sahu-Nature}. In the intermediate temperature range, where single thermally activated phase-slips are responsible for switching events, the standard deviation decreases with cooling \cite{Kurkijarvi-PRB, Clarke-Science}. At the lowest temperatures the standard deviation saturates, indicating that quantum tunneling is the dominant process by which the system escapes from the metastable superconducting state. The temperatures at which the standard deviation saturates are taken as the quantum crossover temperatures, $T_q$, and recorded in Table~\ref{table-params}. We observe that in the normal temperature range in which photon-detectors are operated, T $\approx$ 5 K, multiple phase-slips are responsible for the dark counts. Thus the models which only take into account single phase slips should not be sufficient. 

\begin{figure}[t!]
  \includegraphics[width=9cm]{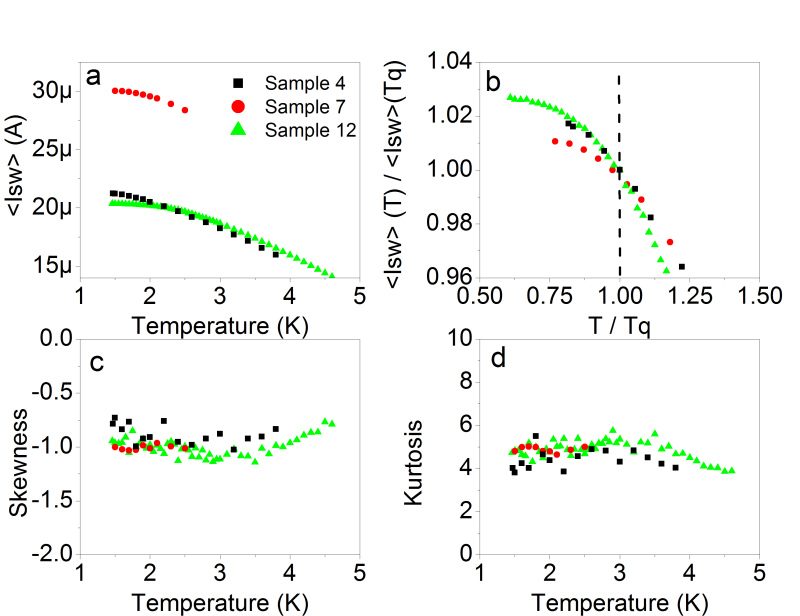}
  \caption{The (a) mean switching current and (b) normalized mean switching current of Samples 4, 7 and 12 increase with cooling in all temperature regions, including $T < T_q$. The normalized mean switching current is obtained by dividing the mean switching current by its value at $T=T_q$ and the normalized temperature is obtained by dividing $T$ by $T_q$. The (c) skewness and (d) kurtosis reflect expected values and verify that the saturation in standard deviation is not caused by reaching a base level of noise in the system. The legend presented in (a) pertains to all four plots in this figure.
 }\label{Fig-stats}
\end{figure}

While the standard deviation saturates at low temperatures, the mean switching current, $<I_{sw}>$, shown in Fig.~\ref{Fig-stats}a, does not. To make this trend clearer we normalize $<I_{sw}>$ by its value at $T_q$, normalize the temperature by $T_q$ and focus on the lowest temperatures in Fig.~\ref{Fig-stats}b. Analyzing the mean switching current provides a check that the saturation seen in the standard deviation is not simply due to a saturation of temperature, because we would expect to see a similar saturation in the mean switching current in this case. In fact, what is observed is that $\sigma$ shows a sharp saturation at $T_q$ while $<I_{sw}>$ does not exhibit any peculiarity at $T_q$ at all. This type of behavior is expected from the Kurkij\"arvi-Garg model \cite{Aref-PRB}.

\begin{figure}[t!]
  \includegraphics[width=9cm]{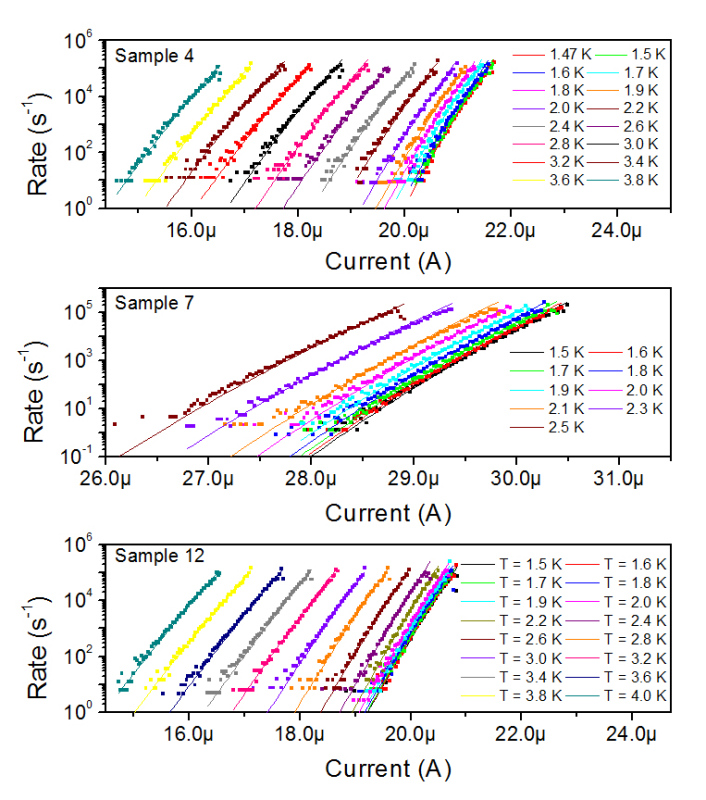}\vskip-.5cm
  \caption{ Rates of phase slips (symbols) for Samples 4, 7 and 12 versus current and fits (solid lines) using the KG model for thermal activation. Fit parameters are shown in Table \ref{table-params} and Fig.~\ref{Fig-tesc}. The lowest temperature curves correspond to the highest currents. At low temperatures, the average value of the switching current changes slowly and the rate curves become more difficult to distinguish in these plots. 
 }\label{Fig-rates}
\end{figure}

The skewness of the switching distributions is about -1 at all temperatures and kurtosis is about 5, as shown in Fig. \ref{Fig-stats}c and \ref{Fig-stats}d. It has been shown that these values are expected for switching distributions in the temperature range where escape is caused by single phase-slips, whether quantum or thermal \cite{Murphy-PRL}. Any perturbations, such as noise in the setup, an influx of external photons, or interference due to cosmic rays would pull skewness toward zero and kurtosis toward three, corresponding to a Gaussian distribution. Insufficient current resolution would have a similar effect. Skewness and kurtosis can also be used to detect the temperatures at which multiple phase-slips must be responsible for escape, as distributions become increasingly Gaussian for $T > T_m$. This trend is most easily seen in Sample 12, which is slightly less noisy than Sample 4. Impacts from high-energy cosmic rays or products of radioactive decay events that can cause dark counts in superconducting kinetic-inductance detectors \cite{Visser-PRL, Swenson-APL} can additionally be ruled out as the source of dark counts in our setup because they result in a much slower dependence of switching rates on bias current than those found in Fig.~\ref{Fig-rates}. The upper limit for the rate of the events with high energy transferred to the detector (greater than of visible-light photon) was set as $10^{-4} s^{-1}$ \cite{Korneev-APL}.  Therefore we conclude that our switching measurements are dominated by thermal or quantum phase-slips.

Switching rates were calculated by performing the Kurkij\"arvi-Fulton-Dunkleberger transformation \cite{Kurkijarvi-PRB, Fulton-PRB} on the switching current distributions, and fit using the Kurkij\"arvi-Garg model \cite{Kurkijarvi-PRB, Garg-PRB, Bezryadin, Aref-PRB}. We assume that the energy barrier for thermal escape can be approximated in the general form of

\begin{equation}
\Delta U = U_0 (1-\frac{I}{I_c})^{b}
\end{equation}

\noindent where $U_0 = \alpha \hbar I_c /2 e$ is the energy barrier at zero current. Here, $I$ is the applied current, $I_c$ is the temperature-dependent critical current and $e$ is the charge of an electron. $\alpha$ and $b$, both positive constants, are used as fitting parameters. This form allows our results to be easily compared to Josephson junctions and long wires. In the LAMH model of long wires \cite{Langer-PR, McCumber-PRB, Tinkham-APL, Tinkham}, $\alpha$ = $\sqrt{6}$ and $b$ =5/4. For Josephson junctions, $\alpha$ = 4$\sqrt{2}$/3 and $b$ = 3/2 \cite{Bezryadin}. At high currents the energy barrier for a phase-slip is expected to become narrow. We can roughly estimate the width of the energy barrier from the energy of a vortex core, $B_C^2 V/2 \mu_0$, and the work done by the Lorentz force on a vortex, $J \phi_0 d x$, where $d$ is the film thickness, $J$ is the supercurrent density, $\phi_0$ is the flux quantum, and $x$ is the distance of the vortex core measured from the edge of the strip. $B_C$ is the critical magnetic field, and V is the volume of the vortex, which we will estimate as $V = \pi \xi^2 d$ where $\xi$ is the coherence length \cite{Tinkham}. The energy barrier reaches zero when the core energy equals the work performed by the Lorentz force at $x_b = B_C^2 \pi \xi^2 / 2 \mu_0 J \phi_0$. Using equation 4.20 from reference \cite{Tinkham}, our estimate can be rewritten as $x_b = \phi_0 w / 16 \pi \Lambda \mu_0 I$. Plugging in a typical value of current, $I = 20$ $\mu$A, we find $x_b \approx$ 4 nm. Physically, $x_b$ represents a rough estimate for the barrier width for a phase-slip event in our quasi-2D superconducting strip. After crossing this energy barrier, the system will develop a vortex at $x_b$ which will then move classically across the strip causing a phase-slip. Because the relevant region for phase-slippage is on the order of the coherence length, our system should be able to be described similarly to a quasi-one-dimensional wire, and we can justify our choice of the barrier form (Eq. 1). While this form is quite similar to that of junctions or wires, it is general enough for the small region of currents examined at any given temperature that fits to this form do not preclude the validity of other models, such as the single-vortex model. However, we note that detailed theory \cite{Vodolazov-PRB} also supports the idea that at high currents the energy saddle point is located near the edge of the film, effectively reducing the problem to a quasi-one-dimensional case.

Using the Arrhenius activation equation, we fit the rates to

\begin{equation}
\Gamma = \Omega \exp(\frac{ -\Delta U}{k_{B} T_{esc}}).
\end{equation}

\noindent $\Gamma$ is the rate of phase-slips, $\Omega$ is the attempt frequency, $T_{esc}$ is the escape temperature, which may or may not equal the sample temperature, and $k_B$ is the Boltzmann constant. 

The critical current was determined at each temperature by following the KG model \cite{Bezryadin, Kurkijarvi-PRB, Garg-PRB} $I_c =< I_{sw}> + K \sigma$ where $K = b \sqrt{6} ln(\Omega \sigma / \nu_I)  / \pi$  and $\nu_I$ is the sweep speed. The only temperature-dependent variables in the equation for $K$ are found in the logarithm, and therefore $K$ is expected to vary slowly. The actual value of $K$ varies slowly with temperature within $\pm$ 1 of the values listed in Table~\ref{table-params}.

For simplicity, we ignore any current or temperature dependence in the attempt frequency. It should be noted that the attempt frequency often appears within a logarithm in our analysis and therefore small changes in $\Omega$ should not have a large effect on our results. Additionally, we have observed that good fits can be obtained with different values of the attempt frequency. We therefore choose to show fits at $\Omega =$ 1 THz, on the order of the natural frequency of the superconducting gap oscillations in NbN \cite{Beck-PRL}.

To find the power $b$, we make use of the second relationship in the KG model \cite{Kurkijarvi-PRB, Garg-PRB, Aref-PRB, Bezryadin}

\begin{equation}
\sigma = \frac{\pi I_c}{\sqrt{6} b} (\frac{U_0}{k_B T})^{-1/b} \ln(\frac{\Omega \sigma}{\nu_I})^{1/b-1}.
\end{equation}

\noindent If we assume that the logarithmic term varies slowly, we can derive the relationship $\sigma \propto I_c^{1-1/b} T^{1/b}$, because $U_0 \propto I_c$. We use this relationship to determine a best-fit value of $b$ for the data corresponding to thermal activation by single phase-slips, i.e. $T_q < T < T_m$. Fits are shown in Fig.~\ref{Fig-sigma}b and best-fit values of $b$ are listed in Table~\ref{table-params}. The best-fit values of $b$ found are of the same order of magnitude as predicted for thin wires and/or Josephson junctions. The reason for this similarity is that the energy saddle point is located near the edge of the strip at high bias currents, thus making the strip in many ways similar to a thin superconducting wire \cite{Vodolazov-PRB}.

\begin{figure}[t!]
  \includegraphics[width=8cm]{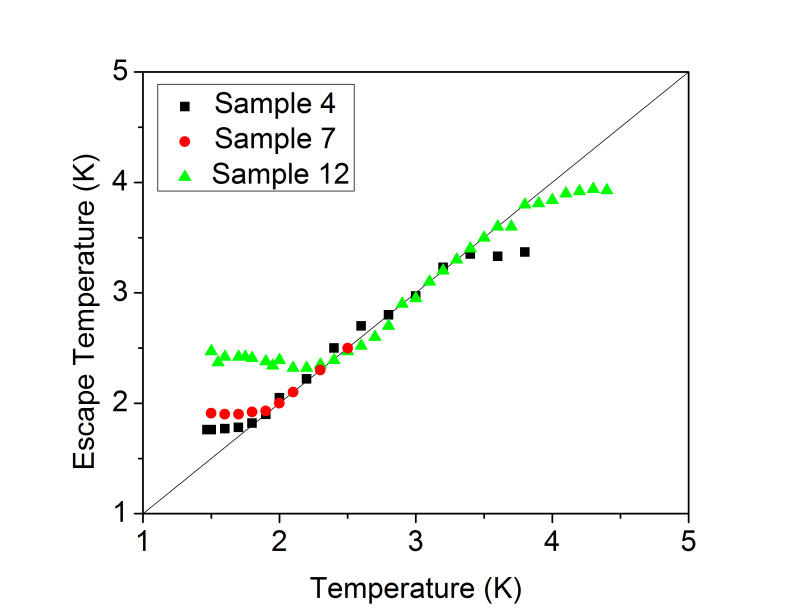}\vskip-.5cm
  \caption{Escape temperature vs real temperature for Samples 4, 7 and 12. The escape temperature was determined by fitting rate curves to Equation 2. Escape temperature was the only fitting parameter allowed to change with temperature for each sample measured. The line $T_{esc} = T$ is plotted for comparison.
 }\label{Fig-tesc}
\end{figure}

After the critical current and the power $b$ have been determined, the constant $\alpha$ is set as a temperature independent parameter and is chosen such that the rates curves calculated by Eq. 2 fit the data well for $T_q < T < T_m$. The escape temperature is then adjusted at all temperatures to produce best-fit curves. The best-fit escape temperature values are plotted versus bath temperature in Figure \ref{Fig-tesc}. The rate curves, shown in Fig.~\ref{Fig-rates}, fit the data well using $T_{esc} \approx T$ for $ T_q < T < T_m$, but at low temperatures best-fits require $T_{esc} > T$. This is a signature of the crossover from thermal activation to quantum tunneling escape processes. The temperatures at which the escape temperature saturates, $T_q'$, are very similar to the values of $T_q$. Values of $T_q'$ are recorded in Table~\ref{table-params}. For all three samples, saturation of escape temperature occurs around 2 K. 

\begin{figure}[t!]
  \includegraphics[width=8cm]{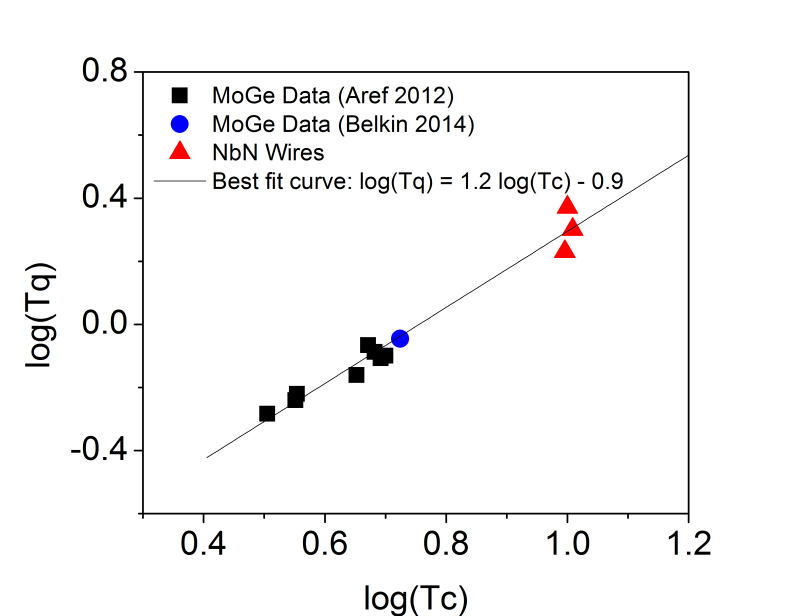}
  \caption{The logarithm of Tq vs the logarithm of Tc plotted with data from MoGe samples \cite{Aref-PRB, Belkin-Arxiv}. There is a general trend that Tq increases with Tc. The best-fit relationship from the combined data is $log(T_q) = 1.2 log(T_c) - 0.9$.
 }\label{Fig-tq}
\end{figure}

Previous work \cite{Aref-PRB} has argued for a linear relationship  between quantum and critical temperatures. In Figure~\ref{Fig-tq} we combine our data for NbN wires with data from MoGe wires and find a best-fit relationship $ T_q \propto T_c^{1.2}$. Thus we confirm the expectation that the relationship is near linear. For this fit, we used the $T_q$ determined from the standard deviation curves of our NbN strips, although using $T_q'$ for this fit would not give a significantly different result. It is important to emphasize that the result shown as a blue circle was obtained using a microwave setup normally used to measure superconducting qubit \cite{Belkin-Arxiv}. Its noise level is low and is qualitatively different compared to the DC measurement setups. Thus the coincidence of the quantum temperatures confirms the conclusions that the observed crossover is not due to any uncontrolled noise but due to the internal quantum fluctuations occurring in the nanowires and thin films.

\begin{table*}[ht]
\caption{Sample parameters and fitting parameters.}
\centering
\begin{tabular}{ c @{\hskip .5cm}  c  @{\hskip .5cm} c @{\hskip .5cm}  c @{\hskip .5cm}  c @{\hskip .5cm} c @{\hskip .5cm} c  @{\hskip .5cm} c  @{\hskip .5cm} c @{\hskip .5cm}  c }
\hline \hline

Sample &  $T_c$ (K) & $T_q$ (K) & $T_q'$ & $b$ & $\alpha$ & $I_{sw}$ at 1.5 K ($\mu$A) & $I_c$ at 1.5 K ($\mu$A) & $R_n$ (M$\Omega$)  &  $K$  \\
[0.5ex]
\hline
4& 9.9 & 1.8 & 1.7 & 1.94 & 1.16 & 21.2 & 26.5 & 1.6 & 27    \\
7 & 10.2 & 1.9 & 1.9 & 2.02 & 1.19 & 30.0 & 36.7 & 1.3 &  28 \\
12 & 10 & 2.4 & 2.3 & 1.36 & 1.08 & 20.4 & 23.9 & 1.9 & 19 \\
[1ex] 
\hline
\end{tabular}
\label{table-params}
\end{table*}

\section{IV. Conclusions}

We have demonstrated that at low temperatures dark counts are dominated by macroscopic quantum tunneling in wide NbN strips biased at currents close to the critical current. By observing the first four moments of the switching distributions, we have checked robustly for noise and other sources of measurement error. We find that quantum tunneling overtakes thermal activation as the dominant process for phase-slips around 2 K in all of our samples, and that the escape rate can be written in the same general form as is done for thin superconducting wires and as well as Josephson junctions. 

\section{Acknowledgements}

A.B. and A.M. acknowledge support from the NSF grants No. ECCS-14-08558 and No. DMR-10-05645, Yu. K. and A.K. acknowledge support from the Ministry of Education and Science of the Russian Federation (State contract No. 14.B25.31.0007), A.S. acknowledges support by grant of the President of the Russian Federation (contract No. M$\Ka$-6184.2014.2) and from the Ministry of Education and Science of the Russian Federation (State demand No. 2327), G.G. acknowledges support by grant of the President of the Russian Federation (contract No. H$\Sh$-1918.2014.2). Authors thank MSPU technology team for fabricating samples and T. M. Klapwijk for discussion of the results.

\end{document}